\documentclass[prd,aps,nofootinbib]{revtex4}
\usepackage{slashed}
\usepackage{graphicx}
\usepackage{amsmath}
\usepackage{amssymb}
\usepackage{amsfonts}
\usepackage{indentfirst}
\usepackage{color}

\newcommand{\be}{\begin{equation}}
\newcommand{\ee}{\end{equation}}
\newcommand{\bea}{\begin{eqnarray}}
\newcommand{\eea}{\end{eqnarray}}

\newcommand{\nn}{\nonumber\\}

\begin{document} 

\preprint[\leftline{KCL-PH-TH/2020-{\bf 16}}

\title{\Large{\bf On the consistency of a non-Hermitian Yukawa interaction}}

\author{Jean Alexandre, Nick E. Mavromatos} 

\vspace{0.1cm}
\affiliation{Theoretical Particle Physics and Cosmology Group, Department of Physics, King's College London, Strand, London WC2R 2LS, UK}

\begin{abstract}

\centerline{{\bf Abstract}}

We study different properties of an anti-Hermitian Yukawa interaction, motivated by a scenario of radiative anomalous generation of  
masses for the right-handed sterile neutrinos. The model, involving either a pseudo-scalar or a scalar, is consistent both at the classical and quantum levels,
and particular attention is given to its properties under improper Lorentz transformations. 
The path integral is consistently defined with a Euclidean signature, and we discuss the energetics of the model, including the options for
dynamical mass generation.

\end{abstract}

\maketitle

\section{Introduction and Motivation}

Contrary to popular belief, quantum mechanical systems with non-Hermitian Hamiltonians have been argued to be consistent, 
in the sense of being characterised by real energy eigenvalues, thus acquiring potential physical significance, 
in the framework of $PT$ symmetry~\cite{Bender}, which by now has found many experimental and theoretical applications.
Although the quantum-mechanical $PT$ symmetric non-Hermitian Hamiltonian are well understood by now,  
the corresponding field theoretical analogues~\cite{BJR} have only recently started becoming the object of systematic 
studies~\cite{BHMS,AB,ABM,AMS,AEMS1,BKB,similarity},  with potential phenomenological relevance to particle physics. 

In general non-Hermitian interactions may not be characterised by $PT$ symmetry. In such a case, their physical significance 
is not as clear as in the $PT$ symmetric case, given that such systems are not always characterised by real energy eigenvalues. 
Nonetheless, there are classes for which there is physical interpretation. 
One such class, which we shall restrict our attention on in this note, is that of non-Hermitian interactions of Yukawa type (NHY). 
NHY interactions are considered in \cite{KK}, where the authors consider complex energies in order to give an effective
description of particle decay, more specifically Higgs decay to a pair of opposite-sign leptons. 

In our note we shall instead consider an anti-Hermitian Yukawa interaction (AHY), not necessarily $PT$-symmetric, but which allows 
a regime where the energies are real. Comments on non-Hermitian Yukawa neutrino sector with real energies can already be found in \cite{ABM}, 
which studies a non-Hermitian extension of Quantum Electrodynamics. 

Such AHY interactions may appear in the scenario of \cite{MP}. Let us explain why. Such a model is inspired by 
string theory considerations, and describes a radiative anomalous generation of (Majorana) masses for the right-handed sterile neutrinos, 
$\psi_R$, as a result of shift-symmetry breaking interactions of the latter with axions of Yukawa type 
\begin{align}\label{arhn}
i \lambda \, a(x) \, \Big(\overline{\psi_R^C} \, \psi_R - \overline \psi \, \psi_R^C \Big) = i \, \lambda\, a(x) \,  \overline{\psi^M} \gamma^5 \psi^M.
\end{align}
where $C$ denotes the charge conjugate, and $\psi^M = \psi_R^C + \psi_R  $ denotes the Majorana field. 
In the model of \cite{MP} the Yukawa couplings $\lambda \in {\tt R}$  are real and the interaction  \eqref{arhn} is Hermitian. 

The model involves a kinetic mixing of the axion field $a(x)$, which could be one (or more) of the axions that characterise string 
theory effective models, stemming from string moduli~\cite{arvanitaki} 
 with a fundamental axion $b(x)$ that, in four (uncompactified) space-time dimensions, stems from the spin-one (Kalb-Ramond) field 
 of the massless (bosonic) gravitational multiplet of the underlying string theory~\cite{strings}: 
\begin{align}\label{mixing}
\gamma \, \partial_\mu b(x) \, \partial^\mu a(x),
\end{align}
where the parameter $
\gamma \in {\tt R}$ has been taken to be real in \cite{MP}. In \cite{MP} it was assumed that the field $b(x)$ couples to global gravitational anomalies. 
Upon diagonalisation of the kinetic terms  \eqref{mixing} by appropriate field redefinitions, the effective action 
contains 
a kinetic term for the axion field $\frac{1}{2} (1-\gamma^2) \partial_\mu a \, \partial^\mu a $, which thus necessitates  
the restriction 
\begin{align}\label{restr}
|\gamma | < 1~, 
\end{align}
to avoid problems with unitarity.

Moreover, the axion couples now to the Gravitational anomaly with a $\gamma$-dependent coupling~\cite{MP}
\begin{align}\label{gan}
 - \frac{\gamma\, c_1\,
        a(x)}{192 \pi^2\, M_{\rm Pl}\, \sqrt{1 - \gamma^2}}
      {R}^{\mu\nu\rho\sigma} \tilde{R}_{\mu\nu\rho\sigma} 
 \end{align}     
 in a standard notation, with $M_{\rm Pl}$ the reduced Planck mass, $R^{\mu\nu\rho\sigma}$  the Riemann tensor and 
 $\tilde{R}_{\mu\nu\rho\sigma} $ its dual, and $c_1 \propto N_f$, a constant depending on the details of the 
 chiral fermion sector of the underlying model, consisting of $N_f$ fermionic species that circulate in the anomalous graviton loop~\cite{anomalies}.

However, if one  is prepared to use non-Hermitian Hamiltonian, with an imaginary mixing term $\gamma = i \tilde \gamma$, with $\tilde \gamma$ real, 
then the restriction \eqref{restr} in the range of $\tilde \gamma$ is lifted, at the cost that, upon canonically normalising the axion-$a(x)$ kinetic term, 
one obtains an effective Yukawa coupling 
\begin{align}\label{arhn2}
\frac{i\lambda}{\sqrt{1 + \tilde \gamma^2}} \, a(x) \, \Big(\overline{\psi_R^C} \, \psi_R - \overline \psi \, \psi_R^C \Big) 
= \frac{i\lambda}{\sqrt{1 + \tilde \gamma^2}}\, a(x) \,  \overline{\psi^M} \gamma^5 \psi^M.
\end{align}
 
 The radiative mechanism of \cite{MP}, entails the generation of a (Majorana) mass for the sterile neutrino
 \begin{align}\label{massR}
 M_R \propto c_1 \frac{\gamma\,  \lambda}{\sqrt{1-\gamma^2}}\, M_{\rm Pl}^{-8} \, \Lambda^9 , 
 \end{align}
 where $\Lambda$ is the Ultra Violet (UV) momentum cutoff used for the regularisation of the effective gravitational field theory~\cite{donoghue}, 
 which could in general be lower than 
 $M_{\rm Pl}$, although in most cases the two scales are identified. Notably, the mass $M_R$ is independent of the details of the axion potential and, 
 thus, the way its mass arises in the model. 
 
We now remark that, in the extended model in which $\gamma = i \tilde \gamma$, $\tilde \gamma \in {\tt R}$, 
one observes that the radiative fermion mass \eqref{massR} can remain real provided one considers AHY interactions \eqref{arhn2} with purely 
imaginary coupling $\lambda = i \tilde \lambda$, where $\tilde \lambda \in {\tt R}$:
\begin{align}\label{arhnNH}
-\frac{\tilde \lambda}{\sqrt{1 + \tilde \gamma^2}} \, a(x) \, \Big(\overline{\psi_R^C} \, \psi_R - \overline \psi \, \psi_R^C \Big) 
= -\frac{\tilde \lambda}{\sqrt{1 + \tilde \gamma^2}}\, a(x) \,  \overline{\psi^M} \gamma^5 \psi^M~.
\end{align}
In such a case the gravitational anomaly term \eqref{gan} would also be anti-Hermitian. Such a term though vanishes for flat (or Robertson-Walker) 
metric backgrounds, which we restrict our attention to in this work.\footnote{On the other hand, in theories where  the fermions couple to gauge fields, 
even if the gravitational anomalies vanish in background space-times, the axions couple to gauge global anomalies. In general, therefore, 
the anomalous terms will be proportional to the non-zero divergence of the axial (chiral) fermionic current. Hence, in the extended model of \cite{MP} 
with anti-Hermitian anomaly terms \eqref{gan}, with $\gamma =i\tilde \gamma, \tilde \gamma \in {\tt R}$, we will have shift-symmetry preserving 
interactions of the form: 
\begin{align}\label{daj5} 
i \, \frac{\tilde \gamma}{\sqrt{1 + \tilde \gamma^2}}\,  a(x)\, \partial_\mu J^{5\mu}~, \quad \tilde \gamma \in {\tt R}~.
\end{align} 
In the current work we ignore the role of such anomalous interactions.}
The point of the current note is to discuss some properties of such AHY interactions and demonstrate the consistency of the approach. 

 The structure of the article is the following: in the  next section \ref{sec:ahm}, we discuss the most important properties 
 of the model with anti-Hermitian Yukawa interactions, and demonstrate its consistency, from the point of view of unitarity and Lorentz covariance, 
 including improper Lorentz transformations. We also point out that the anti-Hermitian Yukawa interactions is $CPT$ even for pseudo-scalar particles, 
 of interest in the model of \cite{MP}. In section \eqref{sec:ener}, we discuss the energetics of the model, demonstrating that dynamical mass 
 generation for the fermions cannot occur in this model for quite generic reasons, unless extra interactions are included. 
 Finally, section \ref{sec:concl} contains our conclusions and some open questions to be studied in the future.

\section{Anti-Hermitian Yukawa model \label{sec:ahm}}

We consider a Dirac fermion and a real scalar field, with the Lagrangian 
\be\label{model}
L=\frac{1}{2}\partial_\mu\phi\partial^\mu\phi-\frac{M^2}{2}\phi^2+\bar\psi i\slashed\partial\psi-m\bar\psi\psi+\lambda\phi\bar\psi\gamma^5\psi~.
\ee
For real $\lambda$, the Yukawa interaction is imaginary, since $\bar\psi\gamma^5\psi$ is anti-Hermitian: $(\bar\psi\gamma^5\psi)^\dagger=-\bar\psi\gamma^5\psi$.
Under discrete transformation we have 
\bea
\mbox{charge conjugation}~~~~C&:&~~~~\psi(t,\vec r)\to\psi'(t,\vec r)=-i\gamma^2\psi^\ast(t,\vec r)\\
\mbox{parity}~~~~P&:&~~~~\psi(t,\vec r)\to\psi'(t,-\vec r)=\gamma^0\psi(t,\vec r)\nn
\mbox{time-reversal}~~~~T&:&~~~~\psi(t,\vec r)\to\psi'(-t,\vec r)=i\gamma^1\gamma^3\psi^\ast(t,\vec r)~,\nonumber
\eea
such that the anti-Hermitian interaction $\phi\bar\psi\gamma^5\psi$ is:
\begin{itemize}
 \item $PT$-even if $\phi$ is a scalar;
 \item $PT$-odd if $\phi$ is a pseudo-scalar;
 \item $C$-odd in any case.
\end{itemize}
Therefore the AHY interaction we consider is $CPT$-even if $\phi$ is a pseudo-scalar, which gives an extra motivation for considering axions.

\subsection{Equations of motion}

The equation of motion for the (pseudo) scalar field is
\be\label{scaleq}
\Box \phi+M^2\phi =\lambda\bar\psi\gamma^5\psi~,
\ee
where the left hand side is real and the right hand side is imaginary. 
Hence we obtain the equation of motion $\Box\phi+M^2\phi=0$ for the scalar field,
and the following on-shell constraint for fermions
\be\label{constraint}
\bar\psi\gamma^5\psi=0~,
\ee
which means that no chiral condensate is allowed. This condition also allows consistent properties under improper Lorentz transformations. 
Indeed, if $\phi$ is a pseudo-scalar, then both left- and right-hand sides of the equation of motion (\ref{scaleq}) are $P$-odd, since $\bar\psi\gamma^5\psi$
is a pseudo-scalar. But if $\phi$ is a scalar, then the only consistent interpretation of eq.(\ref{scaleq}) is to have both left- and right-hand sides
vanishing, in order to respect improper-Lorentz-transformations covariance properties.

The equation of motion for the fermion is more subtle to interprete, due to the anti-Hermitian interaction. 
Indeed, the variation of the action with respect to $\bar\psi$ gives
\be\label{equaferm}
i\slashed\partial\psi-m\psi+\lambda\phi\gamma^5\psi=0~~~~\Leftrightarrow~~~~
i\partial_\mu\bar\psi\gamma^\mu+m\bar\psi+\lambda\phi\bar\psi\gamma^5=0~.
\ee
On the other hand, the variation of the action with respect to $\psi$ leads to
\be
i\partial_\mu\bar\psi\gamma^\mu+m\bar\psi-\lambda\phi\bar\psi\gamma^5=0~,
\ee
showing a mismatch with eq.(\ref{equaferm}) in the sign of the coupling constant $\lambda\to-\lambda$. 
A first consistent set of equations of motion is therefore obtained by the variation with respect to $\bar\psi$, together with the Hermitian 
conjugation of this equation. A second set of consistent equations is obtained by the variations with respect to $\bar\psi$,
together with the Hermitian conjugation of this equation, which feature the opposite sign for $\lambda$. 
These two sets of equations are physically equivalent though, since quantum corrections in this model depend on $\lambda^2$ only, 
as well as physical processes such as fermion-fermion or scalar-scalar scattering.
We note that an alternative approach is possible, based on similarity transformations,
which does not feature the above ambiguity in the derivation of the equations of motion \cite{similarity}.\\

\subsection{Conserved current}

In the presence of a constant background scalar field $\phi_0$, the effective mass term for fermions is
\be
\bar\psi(m-\lambda\phi_0\gamma^5)\psi~,
\ee
and it is shown in \cite{BJR} that the corresponding energies are real as long as
\begin{align}\label{lmconstr}
|\lambda\phi_0|\le|m|,
\end{align}
which is assumed in this section.
From the equations of motion (\ref{equaferm}) one can easily obtain
\bea\label{divergences}
\partial_\mu j^\mu&=&2i\lambda\phi_0~\bar\psi\gamma^5\psi~~~~\mbox{where}~~~~j^\mu\equiv\bar\psi\gamma^\mu\psi\\
\partial_\mu j^{5\mu}&=&2im~\bar\psi\gamma^5\psi~~~~\mbox{where}~~~~j^{5\mu}\equiv\bar\psi\gamma^\mu\gamma^5\psi~,\nonumber
\eea
which shows consistent properties under improper Lorentz transformations if $\phi$ is $P$-odd: 
$\partial_\mu j^\mu$ is then indeed a scalar and $\partial_\mu j^{5\mu}$ is a pseudo-scalar. From eqs.(\ref{divergences}), the conserved probability current is 
\be\label{conservedcurrent}
J^\mu=j^\mu-\lambda\frac{\phi_0}{m}j^{5\mu}~,
\ee
which was already derived in \cite{AB}, requiring unitarity of the theory. The probability is indeed less than one, 
and hence the approach is self consistent, if and only if eq.\eqref{lmconstr} is valid, which, as mentioned above, 
is the same requirement that leads to real energy eigenvalues.  There is a subtle point here, however, which deserves careful discussion.
The vanishing of $\partial_\mu J^\mu$ would imply  
the identity 
\be\label{currentsidentity}
\partial_\mu j^\mu=\lambda\frac{\phi_0}{m}\partial_\mu j^{5\mu}~,
\ee
In both cases, whether $\phi$ is a scalar or a pseudo-scalar, the constant $\phi_0$ is $P$-even, such that the right-hand side of
eq.(\ref{currentsidentity}) is a pseudo-scalar, whereas the left-hand side is a scalar, 
therefore violating covariance properties under improper Lorentz transformations.

Nevertheless, because of the additional constraint (\ref{constraint}), 
one can see from eqs.(\ref{divergences}) that the currents $j^\mu$ and $j^{5\mu}$ 
are {\it individually} conserved, such that conservation of the
current $J^\mu$ is actually consistent with improper Lorentz transformations independent of the parity properties of the field $\phi$. 
Therefore, for the specific model (\ref{model}), because the (pseudo)scalar field is real, the vector and the axial currents are individually conserved, 
independently of the parity of the scalar field. 

We now remark that, in a Hermitian theory, conservation of the vector current would have been a direct consequence of its role as 
the Noether current associated with the global $U(1)$ symmetry of the model. However, Noether's theorem does not apply in a general non-Hermitian
theory, as discussed in \cite{AMS}, but in the present situation, the constraint (\ref{constraint}) leads to the expected conservation law 
$\partial_\mu j^\mu=0$.

We finally comment on chiral fermions. For concreteness, we consider one species right-handed neutrinos, $\psi_R$ with vanishing bare mass $m=0$, 
which is relevant in \cite{MP}. In this case, for which   the anti-Hermitian mass $\lambda\phi_0\gamma^5$ must also vanish if we impose real energies 
({\it cf.} \eqref{lmconstr}), 
there are gravitational anomalies~\cite{anomalies}, 
due to the circulation of the right-handed chiral fermion in the graviton loop~\footnote{Sterile neutrinos do not couple to gauge fields, 
otherwise, {\it i.e.}  if we have other (charged) types of chiral fermions, like left-handed leptons and quarks in the Standard Model sector, 
there are also gauge anomalies.},  leading to the non-conservation of the classical chiral current:
\begin{align}\label{cons5}
 \partial_\mu \Big(  \overline{\psi_R}\, \gamma^\mu  \, \psi_R \Big) = -\frac{1}{192\pi^2} {R}^{\mu\nu\rho\sigma} \tilde{R}_{\mu\nu\rho\sigma} 
\end{align}
As remarked above, such anomalies play a crucial role on the radiative fermion mass generation of the model of \cite{MP}, 
and its non-Hermitian extension, and in the latter case will lead to additional non-Hermitian terms of the form \eqref{daj5}. 
But the question of gravity/gauge anomalies is not covered in this letter, where we focus on the minimal model (\ref{model}).

\subsection{Path integral}

In order to define the path integral without potential problems related to convergence, due to an imaginary interaction, 
we switch here to the Euclidean metric, where this interaction behaves as a pure phase term.
Introducing the sources $j,\bar\eta,\eta$ for the fields $\phi,\psi,\bar\psi$ respectively, we have then
\be
Z_\lambda[j,\bar\eta,\eta]=\int{\cal D}[\phi,\psi,\bar\psi]\exp\left(-S_{Herm}-S_{antiHerm}-S_{sources}\right)~,
\ee
where
\bea\label{defS}
S_{Herm}&=&\int d^4x\left(\frac{1}{2}\partial_\mu\phi\partial^\mu\phi+\frac{M^2}{2}\phi^2+\bar\psi i\slashed\partial\psi+m\bar\psi\psi\right)\\
S_{sources}&=&\int d^4x (j\phi+\bar\eta\psi+\bar\psi\eta)\nn
S_{antiHerm}&=&-\lambda\int d^4x~\phi\bar\psi\gamma^5\psi~=~i\lambda\int d^4x ~\Phi
~~~~\mbox{with}~~~~\Phi\equiv \mbox{sign}(i\bar\psi\gamma^5\psi)|\bar\psi\gamma^5\psi|~.\nonumber
\eea
Because of the anti-Hermitian term in the Lagrangian, one cannot in general cancel $\delta S/\delta\psi$ and $\delta S/\delta\bar\psi$ simultaneously, where 
$S=S_{Herm}+S_{antiHerm}$, which can potentially lead to an ambiguity in defining the saddle point in the path integral \cite{AEMS1}. In our case though, 
because of the on-shell constraint $\bar\psi\gamma^5\psi=0$, the saddle point is uniquely defined by the Hermitian part of the Lagrangian.

We note here few relevant properties of the Hermitian conjugate of the partition function:
\begin{itemize}
\item It is straightforward to see that $Z_\lambda^\ast[j,\eta,\bar\eta]=Z_{-\lambda}[j,\eta,\bar\eta]$. Given the dependence of quantum corrections on 
$\lambda^2$, the partition function is then of the form $Z=\rho(\lambda^2)\exp\Big(i\lambda\theta(\lambda^2)\Big)$;
\item The change of functional variable $\phi\to-\phi$ leads to $Z_\lambda^\ast[j,\eta,\bar\eta]=Z_\lambda[-j,\eta,\bar\eta]$;
\item The change of functional variable $\psi\to\gamma^5\psi$ leads to $Z_\lambda^\ast[j,\eta,\bar\eta]=Z_\lambda[j,\bar\xi,\xi]$, 
 where $\xi\equiv-\gamma^5\eta$ and $\bar\xi=\bar\eta\gamma^5$.
\end{itemize}
Using the above properties, one can check that the background scalar field $\phi_b$ is real for vanishing fermion sources, since
\be
\phi_b^\ast=\frac{-1}{Z_\lambda^\ast[j,0,0]}\frac{\delta}{\delta j}Z_\lambda^\ast[j,0,0]=
\frac{-1}{Z_\lambda[j,0,0]}\frac{\delta}{\delta j}Z_\lambda[j,0,0]=\phi_b~,
\ee
and the condensate $<\bar\psi\psi>$ is real for vanishing scalar source, since
\be
<\bar\psi\psi>^\ast=\frac{1}{Z^\ast_\lambda[0,\eta,\bar\eta]}\frac{\delta^2 Z^\ast_\lambda[0,\eta,\bar\eta]}{\delta\eta\delta\bar\eta}
=\frac{1}{Z_\lambda[0,\eta,\bar\eta]}\frac{\delta^2 Z_\lambda[0,\eta,\bar\eta]}{\delta\eta\delta\bar\eta}
=<\bar\psi\psi>~.
\ee

Next we proceed to discuss the Energetics of the AHY interactions. 

\section{Energetics of the Non-Hermitian Interaction \label{sec:ener}}

In this section we discuss some energetic arguments associated with the AHY interactions, which could be of relevance if one is 
interested in establishing whether such interactions alone can lead to dynamical mass generation, 
as an alternative to the aforementioned radiative mass mechanism \eqref{massR} of \cite{MP}. As we shall see, based on quite generic arguments, 
within the model \eqref{model} dynamical mass for the fermions is {\it not possible}.

\subsection{Fermionic effective theory}

To discuss fermion dynamical mass generation, we assume a non-zero bare (pseudo) scalar mass $M \ne 0$. 
The fermionic effective action $S^{ferm}_{eff}$ is obtained after integrating out the scalar field
\bea
\exp(-S^{ferm}_{eff})&\equiv&\exp\left(-\int_x~\bar\psi i\slashed\partial\psi+m\bar\psi\psi+\bar\eta\psi+\bar\psi\eta\right)
\int{\cal D}[\phi]\exp\left(-\int_x\frac{1}{2}\phi G^{-1}\phi+i\lambda\phi~ \Phi\right)\\
&=&\exp\left(-\int_x~\bar\psi i\slashed\partial\psi+m\bar\psi\psi+\bar\eta\psi+\bar\psi\eta
+\frac{\lambda^2}{2}\Phi G\Phi\right)~,\nonumber
\eea
where $G^{-1}=-\Box+M^2$ and $\Phi$ is defined in eqs.(\ref{defS}). If we neglect higher order derivatives, we have then
\be\label{Sefffermion}
S^{ferm}_{eff}\simeq\int_x~\bar\psi i\slashed\partial\psi+m\bar\psi\psi+\frac{\lambda^2}{2M^2}|\bar\psi\gamma^5\psi|^2~,
\ee
which, because of the original imaginary Yukawa interaction, includes a {\it repulsive} 4-fermion interaction, and thus increases the energy of the system. 
This also implies that, on setting the bare fermion mass to zero, $m=0$, in \eqref{model}, we {\it cannot} have dynamical mass generation for the fermions, 
since the corresponding process is related to the formation of a condensate.

This result can be expected from a general argument \cite{VW}, used to study the energetics of parity violation in gauge theories, such as $QCD$:  
due to the pure phase nature of the AHY interaction ({\it cf.} \eqref{defS}), the fermionic effective action satisfies
\bea
\exp(-S^{ferm}_{eff})&\le&\int{\cal D}[\phi]~
\Big|\exp\left(-\int_x~\bar\psi i\slashed\partial\psi+m\bar\psi\psi+\bar\eta\psi+\bar\psi\eta+\frac{1}{2}\phi G^{-1}\phi
+i\lambda\phi\Phi\right)\Big|\\
&=&\int{\cal D}[\phi]~\exp\left(-\int_x~\bar\psi i\slashed\partial\psi+m\bar\psi\psi+\bar\eta\psi+\bar\psi\eta
+\frac{1}{2}\phi G^{-1}\phi\right)~,\nonumber
\eea
such that the Euclidean $S^{ferm}_{eff}$, which plays the role of {\it vacuum energy functional}, 
is {\it larger} than that for the free theory, and one cannot expect fermion dynamical mass generation,  
in contrast to the usual Hermitian case, where such a dynamical mass lowers the energy of the system 
(in fact, in the Hermitian case, integrating out the massive (pseudo) scalar field would produce an {\it attractive} four fermion term, 
capable of inducing dynamical mass generation). The situation changes, of course, if one has additional {\it attractive}, 
and sufficiently strong, four-fermion interaction terms in the model \eqref{model}, {\it e.g.} of the type
\begin{align}\label{4feff}
 -\frac{1}{2\,f_4^2} \left(\bar\psi\gamma^5 \psi\right)^2~,
\end{align}
where $f_4$ has dimensions of mass. 
A detailed analysis in such a case~\cite{AMSo} confirms the possibility of dynamical fermion and (pseudo)scalar mass generation, 
for sufficiently strong four-fermion couplings of the order of the UV cutoff $\Lambda$, used for the regularisation of the UV divergences 
of the effective theory.

\subsection{Scalar effective theory}

The scalar effective action $S^{scal}_{eff}$ is obtained after integrating out massive fermions 
\be
\exp(-S^{scal}_{eff})\equiv\exp\left(-\int_x\frac{1}{2}\partial_\mu\phi\partial^\mu\phi+\frac{M^2}{2}\phi^2+j\phi\right)
\int{\cal D}[\psi,\bar\psi]\exp\left(-\int_x\bar\psi(i\slashed\partial+m-\lambda\phi\gamma^5)\psi\right)~.\nonumber
\ee
For a constant scalar field configuration $\phi_0$, the effective potential is then
\be
U_{eff}(\phi_0)=\frac{M^2}{2}\phi_0^2-\mbox{Tr}\left\{\ln(\slashed p+m-\lambda\phi_0\gamma^5)\right\}~,
\ee
such that 
\bea\label{dUeff}
\frac{dU_{eff}}{d\phi_0}&=&M^2\phi_0-\mbox{Tr}\left\{\frac{-\lambda\gamma^5}{\slashed p+m-\lambda\phi_0\gamma^5}\right\}\\
&=&M^2\phi_0+\frac{\lambda^2\phi_0}{4\pi^2}
\left(\Lambda^2-(m^2-\lambda^2\phi_0^2)\ln\left(\frac{\Lambda^2+m^2-\lambda^2\phi_0^2}{m^2-\lambda^2\phi_0^2}\right)\right)~,\nonumber
\eea
where $\Lambda$ is the UV cut off, and we note that $(\slashed p)^2=-p^2$ with the Euclidean metric. 
The energies are therefore real for $m^2\ge\lambda^2\phi_0^2$, as expected from \cite{BJR}.
It is interesting to note that, in the limit $\lambda^2\phi_0^2\to m^2$, the effective potential 
consists in a mass term only
\be\label{Ueff}
U_{eff}\to \frac{1}{2}(M^{(1)})^2\phi_0^2~~~~\mbox{with}~~~~(M^{(1)})^2=M^2+\frac{\lambda^2}{4\pi^2}\Lambda^2~.
\ee
If one keeps in mind the anti-Hermitian fermion mass term $\lambda\phi_0\bar\psi\gamma^5\psi$, 
obtained for a constant scalar field configuration in the original action \eqref{model}, the limits  $\lambda\phi_0\to\pm m$
correspond to the so-called {\it exceptional points}, where the number of 
fermionic degrees of freedom is halved \cite{ABM}, since one of the chiralities has a vanishing probability density \cite{AB}.
From the point of view of the scalar effective model, the result (\ref{Ueff}) shows that this limit corresponds to a non-interacting theory: 
quantum corrections suppress all interactions and lead to a free system.
In the regime $m^2<\lambda^2\phi_0^2$, the effective potential features a complex dressed mass and an imaginary self-interaction
\be\label{Ueffcomplex}
U_{eff}=\frac{1}{2}\left[(M^{(1)})^2-i\frac{\lambda^2}{4\pi}m^2\right]\phi_0^2+i\frac{\lambda^4}{16\pi}\phi_0^4+~\mbox{real}~,
\ee
where we note that the imaginary parts of the effective potential are {\it finite}. 
As a consequence of eq.(\ref{Ueffcomplex}), decay rates are generated dynamically in the strong field regime.
We note, on the other hand, that the fermionic effective theory (\ref{Sefffermion}) is always real, as
a consequence of the scalar propagator being independent of the fermion background, whereas the fermion
propagator does depend on the scalar background. Thus the present model cannot predict fermion decay. 

Before closing, we make a last comment on scalar dynamical mass generation.  
From eq.(\ref{Ueff}) it is also clear that dynamical generation of an axion mass is possible, 
since on setting the bare mass to zero, $M =0$, one obtains from \eqref{Ueff} 
\begin{align}\label{dynam}
(M^{(1)})^2_0 = \frac{\lambda^2}{4\pi^2}\Lambda^2 ~, 
\end{align}
and this situation is discussed further in \cite{AMSo}, using a Schwinger-Dyson analysis.

\section{Conclusions \label{sec:concl}}

In this note we studied the consistency, as well as several properties, of a quantum field theory model with anti-Hermitian Yukawa interactions, 
that may appear in a number of occasions with potential physical significance. 
One of the most interesting applications/motivations concerns an extension of the model \cite{MP}, involving interactions of sterile neutrinos with axions. 
Anti-Hermitian Yukawa interactions can still lead to consistent (anomalous) radiative fermion mass generation in that model, but not dynamical 
fermion mass generation, 
due to energetics. 

A final comment we would like to make concerns  the possibility to map the anti-Hermitian model to a Hermitian model, by the change $\lambda\to i\lambda$. 
The possibility to generate a dynamical mass in the latter case, but not in the former, shows that physical predictions are not equivalent. 
More generally, the possibility to map a non-Hermitian model onto a Hermitian model is not a trivial question for an interacting theory \cite{AEM2}.

\section*{Acknowledgements}

The work  of NEM and JA is supported in part by the UK Science and Technology Facilities  research Council (STFC) under the research grants ST/P000258/1 and 
ST/T000759/1. NEM also acknowledges a scientific associateship (``\emph{Doctor Vinculado}'') at IFIC-CSIC-Valencia University, Valencia, Spain.

\end{document}